\documentclass[sigconf]{acmart}

\usepackage{tikz}
\usepackage{amsmath}

\usepackage[english]{babel}
\usepackage{amsmath, marvosym}
\usepackage{hyperref}

\usepackage{babelbib}
\usepackage{color}
\usepackage{pgfplots}
\pgfplotsset{compat=1.18}
\usepackage{import}
\usepackage{pifont}

\newcommand{\cmark}{\color{green}\ding{51}}
\newcommand{\xmark}{\color{red}\ding{55}}

\usepackage{booktabs}
\usepackage{listings}
\usepackage{multirow}
\usepackage{ulem}
\usepackage{graphicx}
\usepackage[graphicx]{realboxes}
\usepackage{subcaption}
\usepackage{latexsym}
\usepackage{comment}
\usepackage{xspace}
\usepackage{relsize}
\usepackage{algorithmicx}
\usepackage{algorithm}
\usepackage{rotating}
\usepackage{multicol}
\usepackage{algpseudocode}
\usepackage{paralist}

\usepackage{minted}

\usepackage{listings}     % Robust code formatting
\usepackage{xcolor}       % For colors
\usepackage{subcaption}   % For subfigures

\usepackage[dvipsnames]{xcolor}
\definecolor{electricmintbg}{HTML}{39FF99} 
\newcommand{\hlt}[1]{\setlength{\fboxsep}{1pt}\colorbox{electricmintbg}{#1}}
\definecolor{electricorangebg}{HTML}{F7662D} 
\newcommand{\hlto}[1]{\setlength{\fboxsep}{1pt}\colorbox{electricorangebg}{#1}}
\definecolor{electricpurplebg}{HTML}{fc68eb} 
\newcommand{\hltp}[1]{\setlength{\fboxsep}{1pt}\colorbox{electricpurplebg}{#1}}

\usetikzlibrary{shapes.geometric, arrows.meta, positioning, fit, calc, backgrounds, shadows}

\definecolor{codebg}{RGB}{245,245,245}
\definecolor{highlighter}{RGB}{255,248,220} % Light yellow
\definecolor{highlightborder}{RGB}{238,221,130}
\definecolor{procBlueFill}{RGB}{225,235,248}
\definecolor{procBlueDraw}{RGB}{90,130,180}
\definecolor{procOrangeFill}{RGB}{255,235,215}
\definecolor{procOrangeDraw}{RGB}{210,140,80}
\definecolor{dbGray}{RGB}{180,190,200}

\newcommand{\sysname}{\textsc{XTRIDE}\xspace}

\graphicspath{{gfx/}}

\AtBeginDocument{%
  }

\setcopyright{acmlicensed}
\copyrightyear{2026}
\acmYear{2026}
\acmDOI{XXXXXXX.XXXXXXX}
\acmConference[PREPRINT of Work accepted at ACM CODASPY '26]{}{}{}
\begin{document}

\title{Practical Type Inference: High-Throughput Recovery of Real-World Structures and Function Signatures}

\author{Lukas Seidel}
\affiliation{%
  \institution{RevEng.AI \& TU Berlin}
  \country{Germany}
  \city{Berlin}
}
\email{lukas.seidel@reveng.ai}
\orcid{0009-0006-8190-0300}
\author{Sam L. Thomas}
\affiliation{%
  \institution{RevEng.AI}
  \country{United Kingdom}
  \city{Birmingham}
}
\author{Konrad Rieck}
\affiliation{%
  \institution{TU Berlin \& BIFOLD}
  \country{Germany}
  \city{Berlin}
}

\renewcommand{\shortauthors}{Lukas Seidel, Sam L. Thomas, and Konrad Rieck}

\begin{abstract}
The recovery of types from stripped binaries is a key to exact decompilation, yet its practical realization suffers.
For composite structures in particular, both layout and semantic fidelity are required to enable end-to-end reconstruction.
Many existing approaches either synthesize layouts or infer names post-hoc, which weakens downstream usability. 
This is further aggravated by an excessive runtime overhead that is especially prohibitive in automated environments. \
We present \sysname, an improved n-gram-based approach that focuses on practicality: highly optimized throughput and actionable confidence scores allow for deployment in automated pipelines.
When compared to the state of the art in struct recovery, our method achieves comparable performance while being between $70$ and $2300\times$ faster.
As our inference is grounded in real-world types, we achieve the highest ratio of fully-correct struct layouts.
With an optimized training regimen, our model outperforms the current state of the art on the DIRT dataset by 5.09 percentage points, achieving 90.15\% type inference accuracy overall.
Furthermore, we show that n-gram-based type prediction generalizes to function signature recovery: 
conducting a case study on embedded firmware, we show that this efficient approach to function similarity can assist in typical reverse engineering tasks.
\end{abstract}


%\keywords{type recovery, binary analysis, artificial intelligence, performance}

\received{02 January 2026}

\maketitle

%!TEX root = ../paper.tex
\section{Introduction}
\label{sec:introduction}

Binary reverse engineering is critical for security applications such as malware analysis, vulnerability detection, general code understanding and automated program analysis.
A core challenge in this field is the recovery of structural information from binaries, especially stripped ones where variable types, names, and data structures are irrevocably lost during compilation.
Recovering this missing type information is essential for producing readable, analyzable decompiler output that can support downstream tasks.

While recent advances in machine learning and static analysis have improved type recovery capabilities, practical deployment remains challenging.
Particularly for user-defined structures, both layout fidelity and semantic information are required.
Many existing approaches either synthesize layouts without recovering meaningful names~\cite{osprey}, or leverage large language models that incur prohibitive runtime overhead~\cite{dirty,resym}.
Furthermore, most type recovery systems do not provide a principled mechanism for estimating prediction confidence, making it difficult to filter unreliable predictions or to integrate type recovery into automated pipelines.

Recent work has explored diverse approaches to this problem.
Traditional constraint-solving methods~\cite{osprey,tie,retyped} rely on costly inter-procedural analysis with substantial computational overhead, often requiring hundreds of seconds per binary.
Machine learning approaches using Transformers~\cite{dirty} or Graph Neural Networks~\cite{tygr} can predict types from training corpora but struggle with user-defined structures not seen during training and lack interpretable confidence measures.
Hybrid systems combining static analysis with large language models (LLMs)~\cite{hyres,resym,typeforge} have achieved improved layout recovery and relationship inference at the cost of inflated inference runtimes, focusing on structs as one of the most interesting class of types to recover.

N-gram-based approaches offer a compelling alternative.
The STRIDE system~\cite{stride} demonstrates that matching local token contexts against a database of decompiled code can achieve competitive accuracy while being orders of magnitude faster than LLM-based methods.
However, STRIDE provides only a heuristic score without practical thresholding and no differentiated analysis of its performance on real types, as commonly used benchmarks are often biased towards primitives~\cite{tygr}.
Consequently, our understanding of the approach in practice is limited by open gaps in training regimen design, type-vocabulary coverage, systematic performance characterization, and calibrated confidence estimation for reliable filtering.
These gaps matter most in unsupervised, pipeline-scale settings, where incorrect type application compounds through decompiler propagation and where per-function latency directly determines whether deployment is feasible.

In this paper, we present \textbf{XTRIDE}, an improved n-gram-based type recovery system that addresses these limitations.
We focus on deployability rather than open-world recovery.
XTRIDE targets environments with recurring real types, such as libraries, firmware stacks, and standard software components, where grounded prediction can be integrated end to end into a decompiler's type system.
Our evaluation on DIRT shows that \sysname achieves 90.15\% general type prediction accuracy, with 98.26\% on in-training functions and 68.66\% on out-of-training functions, statistically significantly improving over the state of the art by 5.09 percentage points.
At the same time, our Rust implementation processes the DIRT test set at 0.04 ms per function, compared to STRIDE's reported 8.2 ms per function and DIRTY's reported 200--8500 ms per function, enabling throughput that is suitable for continuous analysis and large binary corpora.
For function-signature recovery on embedded firmware samples, XTRIDE achieves up to 61.27\% precision for out-of-train binaries, demonstrating practical utility for identifying key function candidates for initial analysis.

We compare \sysname to HyRES~\cite{hyres} and TypeForge~\cite{typeforge} on the recovery of struct types.
While these systems excel at recovering structure relationships and semantic information, they prioritize comprehensive syntactic recovery over speed.
\sysname occupies a different point in the design space by trading exhaustive layout recovery on out-of-distribution types for orders-of-magnitude faster inference while maintaining competitive layout recovery performance and providing fully qualified type names from a closed vocabulary.
This makes \sysname particularly suitable for automated pipelines and large-scale security scanning, where runtime constraints and predictable abstention are first-class concerns.

In summary, we make the following contributions:
\begin{itemize}
    \item \textbf{XTRIDE:} an improved variant of STRIDE with upgraded training regimens, optimized database configurations, and throughput-oriented implementation choices that increase accuracy on DIRT while maintaining inference efficiency.

    \item \textbf{Analysis of Applicability:} We evaluate \sysname's performance on recovering complex struct types from real-world binaries. We compare its ability to identify and recover the layout of structs against the state-of-the-art systems HyRES and TypeForge.

    \item \textbf{Definite Confidence Score:} Introducing a calibratable confidence score for threshold-based filtering, we report specific thresholds and their precision-recall trade-offs to enable reproducible filtering strategies in unsupervised deployment.

    \item \textbf{Function Signature Recovery:} We present an experimental extension that applies n-gram matching to function signatures, evaluated on embedded firmware samples as an auxiliary task.
\end{itemize}

%!TEX root = ../paper.tex
\section{Background}
\label{sec:background}

In this section, we provide the necessary technical background for understanding type recovery in binary reverse engineering.

\subsection{Reverse Engineering}

When a program is compiled to produce a binary, substantial information is lost.
The compilation process discards variable names, variable types, comments, and high-level abstractions such as user-defined struct.
While some information can be preserved through debug symbols (e.g., DWARF), production binaries are typically stripped to reduce size and inhibit understandability.
The process of understanding compiled binaries and recovering lost information from them is called reverse engineering.

Decompilers attempt to programmatically recover this lost information through static analysis.
Basic type recovery in mainstream decompilers like IDA Pro and Ghidra can distinguish primitive types (e.g., \texttt{int}, \texttt{long}, \texttt{float}) with reasonable accuracy.
However, user-defined composite types such as structs remain largely unrecoverable without additional analysis.
Compiler optimizations further complicate recovery by disrupting static patterns, eliminating unused variables, and transforming memory access patterns.
This loss of type information greatly inhibits readability and understandability of decompiled code as illustrated in Figure~\ref{fig:decompilation_comparison}.
A key insight for structure recovery is recognizing field access patterns in assembly code.
When a program accesses a field of a structure, the compiled code typically computes an address using a pattern such as:
\[
\text{address} = \text{base} + \text{index} \times \text{element\_size} + \text{offset}
\]
This pattern, denoted $[A + X \cdot Y + Z]$ in prior work~\cite{hyres}, captures array indexing within structures.
Static analysis systems use these patterns to infer structure layouts by observing consistent offset accesses from a base pointer.
However, incomplete dataflow analysis, indirect accesses, and compiler optimizations can obscure these patterns, leading to missed fields or incorrect layout inference.
Missing names for functions and argument types further complicates the task of understanding the program's behavior.

\begin{figure}[h]
    \centering
    % --- First Subfigure: Raw Decompilation ---
    \begin{subfigure}[b]{\columnwidth}
        \begin{lstlisting}[firstnumber=25]
v5 = *(_DWORD *)(v3 + 124);
v6 = @\hlt{(*(int (\_\_cdecl **)(int, \_DWORD))(*(\_DWORD *)(v5 + 80)}@[...];
v4 = v6;
if ( v6 >= 0 )
{
  @\hlt{(*(void (\_\_cdecl **)(int, \_DWORD))(*(\_DWORD *)(v5 + 80)}@[...];
  v7 = *(_BYTE *)(v3 + 5);
  @\hlto{*(\_BYTE *)(v3 + 5)}@ = 0;
  v4 = sub_284D(0, 0, 5, v7, 0, 0, 0);
  @\hlto{*(\_BYTE *)(v3 + 5)}@ = v7;
  @\hltp{*(void (\_\_cdecl **)(int))(dword\_8100 + 136)(2000)}@;
        \end{lstlisting}
        % \caption{Raw decompilation output before type inference.}
        \label{fig:raw_decompilation}
    \end{subfigure}

    % --- Second Subfigure: Refined Decompilation ---
    \begin{subfigure}[b]{\columnwidth}
        \begin{lstlisting}[firstnumber=25]
ParentIf = UsbBulkTransfer->ParentIf;
v6 = @\hlt{ParentIf->HubApi}@->
  ResetPort(ParentIf, UsbBulkTransfer->ParentPort);
v4 = v6;
if ( v6 >= 0 )
{
  @\hlt{ParentIf->HubApi}@->
    ClearPortChange(ParentIf, ...);
  Address = UsbBulkTransfer->Address;
  @\hlto{UsbBulkTransfer->Address}@ = 0;
  v4 = UsbCtrlRequest(UsbBulkTransfer, EfiUsbNoData, ...);
  @\hlto{UsbBulkTransfer->Address}@ = Address;
  @\hltp{gBS->Stall(2000u)}@;
        \end{lstlisting}
        % \caption{Refined decompilation output after applying type recovery.}
        \label{fig:refined_decompilation}
    \end{subfigure}
    \caption{Motivating example for type recovery on decompiled code. Top: Initial output with opaque offsets and generic function pointers. Bottom: The same code segment after correct structures and types have been applied, revealing meaningful field names and function calls.}
    \label{fig:decompilation_comparison}
\end{figure}

\subsection{Type Recovery}
Early approaches to type recovery utilized dynamic analysis to observe memory access patterns at runtime. 
Systems like Howard~\cite{howard} execute binaries to infer types based on offset consistency and system call arguments. 
However, dynamic methods are difficult to automate in headless environments and struggle to achieve the code coverage necessary for complete binary recovery.
Consequently, static analysis has become the dominant paradigm~\cite{tie, retyped}. 
While newer methods utilize algebraic subtyping to improve scalability~\cite{binsub}, deterministic static analysis faces two significant limitations. 
First, constraint generation is computationally expensive, often prohibiting large-scale analysis. 
Second, even those methods that are able to reconstruct structural layouts (e.g., distinguishing a pointer from an integer) cannot recover nominal semantic information (e.g., resolving a layout to struct \texttt{sockaddr}). 
Without access to a known type vocabulary, the output remains semantically opaque to the analyst.
To address the semantic gap, probabilistic methods have been adopted to approximate the human reverse engineering process. 
Graph Neural Networks (GNNs) can utilize dataflow graphs constructed from intermediate representations (IR) to learn usage patterns~\cite{tygr}.
Most recently, Large Language Models (LLMs) have been applied to treat type recovery as a translation task, predicting variable types and names with high semantic fidelity.

\subsection{Language Models}
Treating decompiled code as a natural language allows for the application of probabilistic modeling to the type inference problem from another angle. 
Under this, variable types are inferred not through rigorous constraint solving, but by recognizing semantic and syntactic patterns in the code surrounding a variable's usage.
Transformers~\cite{attention} currently represent the state of the art in this domain. 
These architectures rely on self-attention mechanisms to weigh the significance of different tokens within a sequence, allowing them to capture long-range dependencies and global context. 
In the context of type recovery, Transformer-based models predict types by analyzing the entire function body, effectively translating untyped code into typed code.
While highly accurate, this process requires dedicated hardware and results in significant computational cost. 
N-gram models, a precursor to modern deep learning in natural language processing, offer a lightweight alternative. 
An n-gram is a contiguous sequence of $n$ items from a given text. 
In type recovery, this approach models a variable's type based on its immediate local context~\cite{stride}.
Fundamentally, both architectures leverage the distributional hypothesis: the assumption that a token's meaning (or a variable's type) can be derived from the context in which it appears.
As these approaches model the problem as a classification task, they are limited to the vocabulary of the training data.
Hence, they are applicable to inferring types from common libraries but cannot generalize as well as static analysis approaches.
However, they come with the benefit of adding not only type layout information but also real-world grounded names.
%!TEX root = ../paper.tex
\section{Motivation}
\label{sec:motivation}
We motivate the need for XTRIDE by examining limitations in existing type recovery systems, particularly around confidence calibration, runtime efficiency, and practical deployment.
\begin{table}[t]
    % \footnotesize
    \caption{Survey of type recovery systems. Overhead refers to runtime cost. Struct Layouts indicates whether the system recovers full struct member information.}
    \label{tab:relwork-comparison}
    \small
    \begin{tabular}{l|cccc}
    \toprule
    \textbf{System} & \textbf{Approach} & \textbf{Dataset} & \textbf{Overhead} & \textbf{Struct} \\
     & & & & \textbf{Layouts} \\
    \midrule
    IDA~\cite{ida} & Static Analysis & - & Low & Limited \\
    \& Ghidra~\cite{ghidra} & + Heuristics &  &  & \\
    \midrule
    OSPREY~\cite{osprey} & Constraint & Coreutils & Very & \cmark \\
     &  &  & High &  \\
    \midrule
    TypeMiner~\cite{typeminer} & Classifier & Coreutils & Medium & \xmark \\
    & (RF + SVM) &  & & \\
    \midrule
    DIRTY~\cite{dirty} & Transformer & DIRT & High & \cmark \\
     &  &  &  & (in-train) \\
    \midrule
    TYGR~\cite{tygr} & GNN & TYDA & Medium & \xmark \\
    \midrule
    STRIDE~\cite{stride} & N-gram & DIRT & Low & \cmark  \\
     &  &  &  & (in-train) \\
    \midrule
    HyRES~\cite{hyres} & Static Analysis & Coreutils & High & \cmark \\
     & + LLM &  &  & \\
    \midrule
    ReSym~\cite{resym} & LLM + & Custom & High & \cmark \\
     & Static Analysis &  &  &  \\
    \midrule
    TypeForge~\cite{typeforge} & Constraint & Coreutils & Medium & \cmark \\
     & + LLM &  &  &  \\
    \midrule
    ByteTR~\cite{bytetr} & GNN & TYDA & Medium & \xmark \\
    % \textbf{XTRIDE} & N-gram & DIRT & Low & \cmark \\
    \bottomrule
    \end{tabular}
\end{table}

\textbf{Runtime efficiency for large-scale analysis.}
Many modern type recovery systems rely on large language models or complex static analysis.
DIRTY requires over one second per function on GPU and 8+ seconds on CPU~\cite{dirty}.
For an average UEFI firmware image with 300-500 modules and 50,000+ functions, this translates to multiple days of processing time on CPU and half a day on GPU.
Such overhead is prohibitive for large-scale security scanning platforms or continuous integration pipelines.
Constraint-solving approaches like OSPREY incur similarly high costs, averaging 528.24 seconds per binary on the Coreutils benchmark~\cite{osprey}.
While these methods achieve high accuracy on layout recovery, their computational requirements limit deployment at scale.
N-gram-based approaches offer a compelling alternative.
STRIDE demonstrates that matching decompiled code tokens against a database can achieve 85\% overall accuracy on DIRT with only 8ms per function in Python~\cite{stride}.
While this is a significant improvement over other approaches, it still leaves substantial cost at pipeline scale.
A clearer focus on performance is needed to enable new use cases such as security platforms integrating type recovery into every decompilation request or continuous integration pipelines running type inference every commit.

\textbf{Identifying real types.}
Early type recovery systems~\cite{osprey, debin} only provided primitive type labels such as \texttt{int}, \texttt{char}, \texttt{pointer} and \texttt{struct} but stopped to produce predictions for the layout and member types of a struct.
This severely limits their practical usefulness as structs are of special interest. 
Correctly recovering the layouts and semantics of these composite types greatly facilitates the readability of decompiled code~\cite{R2I}.
Those systems that attempt to recover struct members are still prone to error as they usually recover each field individually, often failing to recover the layout entirely correctly. 
Approaches that treat type recovery as a multi-label classification task~\cite{dirty, stride} avoid this partially, mapping directly to a vocabulary grounded in real-world information, providing full type qualification.
While probabilistic approaches can also attempt to recover layouts of never-before-seen types, vocabulary-grounded prediction is inherently unable to generalize to types not seen during training. 
This can be partially addressed by larger training corpora, e.g, by training on common standard and open source libraries or fine-tuning on context-specific code bases, but will always remain a limitation.
The closed-dictionary approach also provides full semantic information, yielding all type and field names upon prediction. 
Working on fully-named, real data types further increases comprehensibility of decompiled binaries.
Many earlier systems working at field-level lack this kind of information completely.
Those that provide a separate mechanism for name recovery introduce significant additional overhead for this task~\cite{hyres}.
Identifying not only individual variable types but also functions with their parameter types can further enhance decompilation output.

% Function signature-, or even name-, recovery is a powerful tool, providing invaluable information.
% The state-of-the-art systems in function similarity are quite involved~\cite{binaryai}, while many tasks could already profit from simpler estimations with lower fidelity.
% To the best of our knowledge, there is also no unified approach for variable type and function signature recovery.
% Especially in large binaries with thousands of functions, a leightweight approach that not only enhances the decompiler output with types but also can identify a subset of highly interesting functions could save a researcher significant manual effort, even at moderate precision.

\textbf{The need for calibrated confidence.}
Existing type recovery systems produce predictions that indicate relative ranking among candidates rather than a true estimate of correctness.
A user seeking to filter unreliable predictions has no principled way to select a threshold.
This limitation is particularly acute in automated pipelines and a common oversight in research where the goal is evaluation as opposed to real-world applicability.
However, if the type inference system produces many false positives, it becomes counterproductive.
A calibrated confidence score gives the user a tool to trade coverage for reliability, producing fewer but more reliable predictions.
This is particularly important once recovered types are applied to a decompiler and propagated across functions, where an incorrect struct with wrong offsets can cause compounding downstream errors.

\medskip
We survey the field of type recovery systems in Table~\ref{tab:relwork-comparison} and observe that practically all available systems either do not support the full layout recovery of structs or introduce an overwhelmingly large overhead.
STRIDE demonstrated that n-grams can offer an alternative to more elaborate machine learning or constraint solving techniques.
We see several limitations that still hinder practical deployment:
\begin{description}
\item\textbf{[L1] Lack of confidence scoring.}
STRIDE's scoring mechanism aggregates n-gram matches and diversity factors to produce a heuristic score.
This score is useful for ranking candidates but does not correspond to a probability of correctness.
Users cannot reliably filter predictions by threshold.
\item\textbf{[L2] Primitive type bias.}
Primitive types such as \texttt{int32\_t} and \texttt{uint64\_t} dominate training data, comprising over 30\% of all variables.
When a struct type and a primitive type have similar scores, STRIDE's tie-breaking by frequency favors the primitive, potentially leading to missed struct predictions.
A thorough investigation of the model's performance on non-primitive / -trivial types is currently missing.
\item\textbf{[L3] Limited ablations on training configuration.}
STRIDE evaluates on DIRT but does not report detailed ablations on training set size, n-gram database composition, or memory-accuracy trade-offs.
\item\textbf{[L4] Generalization beyond training data.}
N-gram approaches match tokens verbatim, limiting generalization to code patterns not seen during training.
Understanding how to improve downstream accuracy through better training regimens is critical.
At the same time, these systems should be evaluated honestly as closed-vocabulary methods: they are best suited for partially known environments with repeated real types, not for open-world recovery on arbitrarily shifted distributions.
\end{description}
% TOKEN: aqHSs_Y5JSusGDwb06j5V-ldvwszdP0qObZBnof0vLY

% \subsection{Design Goals for XTRIDE}

% Based on these observations, we establish the following design goals for XTRIDE:

% \begin{enumerate}
%     \item \textbf{Confidence score:} Provide a definite confidence score, enabling threshold-based filtering with predictable precision-recall trade-offs.

%     \item \textbf{Efficiency:} Further improve STRIDE's efficiency advantage over LLM-based methods by replacing the Python implementation with a Rust one, enabling deployment in latency-sensitive and large-scale scenarios.

%     \item \textbf{Improved accuracy:} Increase overall accuracy and particularly struct accuracy through optimized training regimens, larger datasets, and better database configurations.

%     \item \textbf{Generalization:} Improve out-of-training accuracy through dataset expansion and address primitive type bias to improve struct type recovery.

%     \item \textbf{Reproducibility:} Provide detailed ablations on training set size, database composition, and confidence thresholds to enable reproducible deployment strategies.

%     \item \textbf{Extensibility:} Demonstrate that n-gram matching can extend beyond variable type recovery to auxiliary tasks such as function-name recovery.
% \end{enumerate}

%!TEX root = ../paper.tex
\section{Design}
\label{sec:design}
In this section, we detail the key architectural choices and workflow of XTRIDE.

\begin{table}[t]
    \centering
    \small
    \caption{Key differences between STRIDE and XTRIDE.}
    \begin{tabular}{lcc}
    \toprule
    \textbf{Feature} & \textbf{STRIDE} & \textbf{XTRIDE} \\
    \midrule
    Actionable Confidence & \xmark & \cmark \\
    Function signature recovery & \xmark & \cmark \\
    Training optimizations & Limited & Extensive \\
    Struct Recovery Evaluation & Limited & \cmark \\
    Throughput Optimization & \xmark & \cmark \\
    \bottomrule
    \end{tabular}
    \label{tab:xtride-vs-stride}
\end{table}

\subsection{Overview}

XTRIDE follows STRIDE's core approach: treating decompiled code as text and matching variable contexts against a database of known patterns.
For each variable occurrence in a function, XTRIDE extracts $n$ tokens to the left and n tokens to the right, forming a context window.
This context is matched against n-gram databases of varying sizes, e.g., $n \in \{2, 3, 4, 5, 6, 7, 8, 9, 10, 11, 12, 13, 14, 15, 30, 60\}$ for the original STRIDE configuration.
These blended n-grams were built from a training corpus of decompiled code with groundtruth type annotations.
Before hashing, tokens are normalized to reduce sparsity: literals are replaced with placeholders and call contexts include the full extracted argument-signature tokens used at the call site.
Throughout this paper, a \emph{match} denotes hash equality on these normalized token sequences.

XTRIDE extends this foundation in three ways:
\begin{enumerate}
    \item \textbf{Improved training:} Larger training sets, optimized database configurations, and separate databases for 32-bit and 64-bit binaries to reduce false positives.
    \item \textbf{Definite confidence score:} Calibrated confidence estimation using held-out validation data, enabling threshold-based filtering.
    \item \textbf{Function signature recovery:} Extension of n-gram matching to function signatures for auxiliary reverse-engineering tasks.
\end{enumerate}

\sysname retains STRIDE's core efficiency and n-gram matching approach while adding calibration, training optimizations, and extensibility to auxiliary tasks, all while reducing inference cost even further.
These enhancements enable practical deployment in automated pipelines and interactive workflows.
Table~\ref{tab:xtride-vs-stride} summarizes key differences between STRIDE and \sysname.

\subsection{N-Gram Database Construction}

XTRIDE constructs n-gram databases from a training corpus of binaries with debug information as groundtruth labels.
Each binary in the training set is first decompiled using a standard decompiler (IDA Pro in our case) with debug symbols to obtain groundtruth types and then tokenized into sequences of tokens.
For each variable occurrence, context windows of varying sizes $n$ are extracted and the mapping from (n-gram hash, type) pairs along with occurrence counts are recorded.
This process is repeated for all training binaries to build a global database.
A candidate type is in the top-$k$ set of a query if it appears among the $k$ highest-scoring labels returned for the queried context after local ranking. Unless stated otherwise, we use $k=3$.
A critical design decision is database granularity.
STRIDE uses 16 separate databases for different values of $n$.
XTRIDE explores configurations with fewer databases (cf. Section ~\ref{exp:training}) to reduce memory overhead while maintaining accuracy.
Our evaluation shows that 4 databases (n $\in \{2, 8, 16, 64\}$) achieve competitive accuracy at the same memory requirements when trained on more samples.

\textbf{Bitness separation.}
For \sysname, we maintain separate databases for 32-bit and 64-bit binaries where possible.
Structure access patterns differ between architectures due to pointer size and alignment differences.
By separating databases, we reduce false positives where a struct access pattern in 32-bit code matches a different struct in 64-bit code.
This design choice is informed by the observation that binary type databases are most effective when fully qualified types resolve to exactly one type per bitness.

\subsection{Improved Scoring and Aggregation}
\label{scoring}

XTRIDE inherits STRIDE's scoring mechanism to rank local candidates but introduces a refined confidence for global ranking and interpretation.
For each variable occurrence, XTRIDE queries databases for multiple values of n and aggregates matches using three factors:
Matches are aggregated by size, where larger n-grams receive higher weights as they capture more specific contexts, frequency, where multiple occurrences of a variable matching the same type increase confidence, and diversity, where n-grams that consistently appear with a single type are more distinctive than those appearing with many types.
With $w_i$ being the weight per n-gram size, $f_i$ the frequency of occurrence, and $d_i$ the diversity factor (inverse of the number of types associated with the n-gram), the raw score $s$ for a candidate type $t$ is:
\[
s(t) = \sum_{i=1}^{k} w_i \cdot f_i \cdot d_i
\]
% The weights $w_i$ are fixed by database size and may optionally be tuned on validation data to better align confidence with observed accuracy.
% To address primitive type bias, XTRIDE implements an optional heuristic: when the top-ranked prediction is a primitive type and a struct type ranks within the top-k candidates (default k=3) with a score ratio above a threshold (e.g., 0.8), XTRIDE promotes the struct type.
Furthermore, XTRIDE introduces a confidence score to enable threshold-based filtering of inference results, addressing L1.
The goal is to transform the raw aggregated score into a calibrated probability that accurately reflects the likelihood of correctness.
We normalize the aggregation scores as follows:
\[
\text{c}_{\text{norm}} =
\begin{cases}
\dfrac{s^{*} - B}{M - B}, & \text{if } M > 0 \wedge M > B \wedge s^{*} > B,\\[6pt]
0, & \text{otherwise,}
\end{cases}
\qquad B=\tfrac{1}{2}M
\]
where $s^{*}$ is highest raw score among candidates, $M$ is the context-based maximum possible score, i.e., the number of n‑gram contexts in which the candidate appears in the database’s top‑k results, and $B$ is the baseline threshold given our per‑context score range $[0.5, 1.0]$.
For each prediction in the validation set, we record the raw score and whether the prediction was correct.
Using isotonic regression~\cite{isotonic}, we can fit a non-decreasing piecewise-constant function that minimizes squared error between predicted probabilities and observed frequencies, mapping raw scores to calibrated probabilities.
This non-parametric approach captures non-linear relationships between raw scores and correctness.
% The resulting definite score is a deploy-time quantity: if the deployment distribution shifts, recalibration on a small representative slice is appropriate; if the distribution remains stable, the calibration transfers directly.

\textbf{Threshold selection.}
A well-calibrated score satisfies the property that among all predictions with confidence $c$, approximately $c$ fraction are correct.
With this property, users can select a threshold $\tau$ to achieve desired precision-recall trade-offs.
We explore performance at different thresholds (0.90, 0.65, 0.40, none) to illustrate the trade-off space.
For production deployment, the threshold can be chosen based on application requirements: high-precision tasks may use $\tau = 0.9$, while exploratory analysis may use $\tau = 0.65$.
If the target test distribution is known in advance, full calibration on an in-domain validation set is recommended.

\subsection{Function Signature Recovery}
\label{funcrec}

XTRIDE extends n-gram matching to function signature recovery as an auxiliary task.
The core insight is that function signatures and calling contexts exhibit similar local patterns to variable usage.
We treat function calls like variable names and extract the corresponding context tokens.
These contexts are matched against a database of function call contexts with full signatures, i.e., including names and types for the function's parameters, as their type labels built from the training corpus.
To avoid type confusion, vocabularies for function signatures and usual types are strictly separated.
The matching and scoring procedure mirrors variable type recovery in the first stage.
As we have global identifiers for functions, e.g., based on their address, we introduce a second aggregation stage: 
for every function address, we aggregate all predictions across function contexts and weigh the inferred best signature match by number of evaluated contexts per function.
This produces a single binary-wide prediction. 
Figure~\ref{fig:pipeline_compact} illustrates the process.

We do not aim to compete with full embedding-based function similarity approaches but present this technique as a fast, low-cost alternative in lower-stake environments.
To the best of our knowledge, this is the first technique to attempt function signature recovery solely based on syntactic analysis of the caller context.

\begin{figure}[t]
    \centering
    \resizebox{0.8\columnwidth}{!}{%
    \begin{tikzpicture}[
        % increased vertical spacing to preventing overlap
        node distance=0.55cm and 0.25cm, 
        font=\sffamily\fontsize{6}{7}\selectfont,
        >=Stealth,
        % --- Styles ---
        basebox/.style={
            draw, rounded corners=2pt, align=center, line width=0.5pt, inner sep=3pt, fill=white
        },
        codebox/.style={
            basebox, fill=gray!5, draw=gray!40, font=\ttfamily\fontsize{5}{6}\selectfont, align=left
        },
        blueproc/.style={
            basebox, fill=blue!5, draw=blue!40, text width=1.9cm
        },
        orangeproc/.style={
            basebox, fill=orange!10, draw=orange!40, text width=2.3cm
        },
        greenproc/.style={
            basebox, fill=green!10, draw=green!40, text width=1.9cm
        },
        arrow/.style={
            ->, draw=black!60, line width=0.6pt, rounded corners=3pt
        },
        arrow_orange/.style={
            ->, draw=orange!80!black, line width=0.6pt, rounded corners=3pt
        },
        arrow_green/.style={
            ->, draw=green!60!black, line width=0.6pt, rounded corners=3pt
        }
    ]
        % --- 1. TOP: CODE ---
        \node[codebox, text width=4.8cm, align=center] (topcode) {
            \textbf{DECOMPILED BINARY}\\
            ... sub\_aaa(a,b); \colorbox{blue!10}{sub\_fff(c);} \colorbox{yellow!20}{sub\_aaa(d,e);} ...
        };
        % --- 2. CALL SITES ---
        % Center Site (Site 2)
        \node[blueproc, below=0.5cm of topcode] (ext2) {
            \textbf{Call Site 2}\\
            N-gram Context: \\
            \texttt{[;, \colorbox{blue!10}{sub\_fff}, (]}
        };
        
        % Left Site (Site 1)
        \node[blueproc, left=0.15cm of ext2] (ext1) {
            \textbf{Call Site 1}\\
            N-gram Context: \\
            \texttt{[x, =, \colorbox{yellow!20}{sub\_aaa}]}
        };
        % Right Site (Site 3)
        \node[blueproc, right=0.15cm of ext2] (ext3) {
            \textbf{Call Site 3}\\
            N-gram Context: \\
            \texttt{[\colorbox{yellow!20}{sub\_aaa}, (, d]}
        };
        % Arrows Code -> Sites
        \draw[arrow] (topcode.south) -- ++(0,-0.15) -| (ext1.north);
        \draw[arrow] (topcode.south) -- (ext2.north);
        \draw[arrow] (topcode.south) -- ++(0,-0.15) -| (ext3.north);
        % --- 3. SCORES & DB ---
        % Increased distance here (0.7cm) to fit the DB icon cleanly
        \node[blueproc, below=0.7cm of ext1] (cand1) {Local Scores\\ \tiny \{SigA:0.9, SigB:0.2\}};
        \node[blueproc, below=0.7cm of ext2] (cand2) {Local Scores\\ \tiny \{SigC:0.4, SigD:0.1\}};
        \node[blueproc, below=0.7cm of ext3] (cand3) {Local Scores\\ \tiny \{SigA:0.8, SigB:0.3\}};
        % MINIATURE DB ICON: Floating in the center gap
        \node[cylinder, shape border rotate=90, aspect=0.25, draw=gray!80, fill=white, 
              minimum height=0.5cm, minimum width=0.4cm, scale=0.7, inner sep=0pt,
              label={[font=\tiny, black, yshift=-0.1cm]right:
                \parbox{2.05cm}{Function Signature\\N-gram DB}
              }] 
              (db) at ($(ext2.south)!0.4!(cand2.north)$) {};
        
        % Interaction Dashes (Subtle)
        \draw[dotted, thick, gray!60] (db.west) -- ++(-0.5,0);
        \draw[dotted, thick, gray!60] (db.east) -- ++(0.5,0);
        % Arrows Sites -> Scores
        \draw[arrow] (ext1) -- (cand1);
        \draw[arrow] (ext2) -- (cand2);
        \draw[arrow] (ext3) -- (cand3);
        
        % --- 4. LOCAL PREDICTIONS ---
        \node[blueproc, below=0.5cm of cand1] (pred1) {Pred: \textbf{sigA}\\ \tiny conf: 0.9};
        \node[blueproc, below=0.5cm of cand2] (pred2) {Pred: \textbf{sigC}\\ \tiny conf: 0.4};
        \node[blueproc, below=0.5cm of cand3] (pred3) {Pred: \textbf{sigA}\\ \tiny conf: 0.8};
        
        \draw[arrow] (cand1) -- (pred1);
        \draw[arrow] (cand2) -- (pred2);
        \draw[arrow] (cand3) -- (pred3);
        
        % --- 5. CONFIDENCE FILTERING ---
        \node[basebox, draw=black!70, below=0.5cm of pred1, fill=gray!5] (calib1) {$c > \tau$ ?};
        \node[basebox, draw=black!70, below=0.5cm of pred2, fill=gray!5] (calib2) {$c > \tau$ ?};
        \node[basebox, draw=black!70, below=0.5cm of pred3, fill=gray!5] (calib3) {$c > \tau$ ?};
        
        \draw[arrow] (pred1) -- (calib1);
        \draw[arrow] (pred2) -- (calib2);
        \draw[arrow] (pred3) -- (calib3);
        
        % Filter results
        \node[greenproc, below=0.4cm of calib1] (pass1) {\textcolor{green!40!black}{\checkmark} sigA};
        \node[below=0.4cm of calib2, align=center, text=red!80!black] (fail2) {\textbf{$\times$}};
        \node[greenproc, below=0.4cm of calib3] (pass3) {\textcolor{green!40!black}{\checkmark} sigA};
        
        \draw[arrow_green] (calib1) -- (pass1);
        \draw[arrow, draw=red!60] (calib2) -- (fail2);
        \draw[arrow_green] (calib3) -- (pass3);
        
        % --- 6. AGGREGATION ---
        \node[orangeproc, below=0.7cm of fail2, text width=3.5cm] (agg) {
            \textbf{GLOBAL AGGREGATION}\\
            \tiny Merge high-conf predictions\\
            \tiny $\sum(\text{Scores}) \times \text{Weight}$
        };
        
        % Background box
        \begin{scope}[on background layer]
            \node[fit=(agg), fill=orange!20, draw=orange!30, rounded corners, inner sep=4pt] (agg_bg) {};
        \end{scope}
        
        % Arrows to Aggregation
        \draw[arrow_green] (pass1.south) -- ++(0,-0.15) -| (agg.north);
        \draw[arrow_green] (pass3.south) -- ++(0,-0.15) -| (agg.north);
        
        % --- 7. OUTPUT ---
        \node[orangeproc, below=0.5cm of agg, text width=3.5cm] (output) {
            \textbf{Final Predictions}\\
            \texttt{Func \texttt{sub\_aaa} $\leftarrow$ sigA}
        };
        
        \draw[arrow_orange] (agg.south) -- (output.north);

    \end{tikzpicture}
    }
    \caption{Overview of function signature prediction and global aggregation procedure. Local predictions are filtered by confidence threshold before aggregation.}
    \label{fig:pipeline_compact}
\end{figure}

\subsection{Inference Procedure}

At inference time, \sysname tokenizes a decompiled function and, for each variable, extracts contextual n-grams from all occurrences to subsequently query the corresponding n-gram databases. 
The evidence from these queries is aggregated into a raw score per candidate type, and candidates are ranked accordingly. 
Raw scores are then calibrated into confidence values; the top-ranked type is emitted only if its confidence exceeds a threshold $\tau$, otherwise \sysname abstains. 
When enabled, a lightweight prioritization heuristic biases the final decision toward struct types. 
The output is the selected type together with its confidence score.

%!TEX root = ../paper.tex
\section{Implementation}
\label{sec:implementation}

XTRIDE is implemented in Rust for efficiency and memory safety.
The core n-gram matching engine uses hash maps for $O(1)$ lookup and supports concurrent queries for parallelization.
The implementation comprises approximately 5,000 lines of Rust code.

\subsection{Feature Extraction and Storage}

For each variable and function call occurrence in a decompiled function, we extract the token context consisting of $n$ tokens to the left and $n$ tokens to the right with $n$ varying per database. For calls, we always extract the whole signature including passed arguments, not counting towards the right $n$.
Tokens are normalized to handle variations in whitespace and formatting.
Literals (strings, numbers) are replaced with generic placeholders (e.g., \texttt{<STRING>}, \texttt{<NUM>}) to reduce sparsity.

\subsection{Training Configuration}

An optimized training procedure is one of the contributions of this work and is explained and evaluated in detail in Section~\ref{exp:training}, addressing L3 and L4.
The configuration of the n-gram database is one of the key performance factors. 
The original STRIDE implementation used 16 databases with sizes \{2, 3, 4, 5, 6, 7, 8, 9, 10, 11, 12, 13, 14, 15, 30, 60\} without elaborating on this setup.
Intuitively, a higher number of smaller context databases allow to match on more local n-grams, improving prediction confidence by having multiple matches.
Matching on large contexts, such as 30 or 60 tokens at a time, is rarer but directly leads to high-confidence recognition, basically indicating that an equivalent pattern was seen during training, with low false positive rate. 
These contexts are stored in separate databases, mapped to their respective type labels.
We serialize n-gram databases using \texttt{rkyv}~\cite{rkyv} to enable fast loading and memory-mapped access.
A database ensemble for 64-bit binaries with 300k training samples and 4 n-gram sizes occupies approximately 15 GB on disk and loads in ca. 560ms, where a larger configuration with 16 databases and a total of 48 GB on disk loads in ca. 640ms, averaged across 10 runs.

\subsection{Runtime Performance}

Our Rust implementation is heavily optimized for performance, profiting from Rust's inherently strong optimizations and speed and making use of multi-processing wherever possible.
Memory footprint is minimized at runtime by using memory-mapped I/O for the multi-gigabyte n-gram databases.

%!TEX root = ../paper.tex

\section{Experiments}
\label{sec:evaluation}

We conducted experiments to answer the following questions:
\begin{itemize}[noitemsep]
    \item[\textbf{[RQ1]}] Can we optimize the training procedure of an n-gram based technique to further improve general type inference performance while maintaining inference speed and memory requirements?
    \item[\textbf{[RQ2]}] How does the performance of \sysname compare with the state of the art in recovering complex struct types?
    \item[\textbf{[RQ3]}] Can a globally comparable confidence score be helpful in steering the model?
    \item[\textbf{[RQ4]}] Can \sysname's function signature recovery aid in typical reverse engineering tasks?
\end{itemize}

\subsection{Experimental Setup}
\label{setup}

\textbf{Datasets.}
For general type inference, we use the DIRT dataset introduced by the original DIRTY publication~\cite{dirty} as the primary benchmark.
It is widely adapted~\cite{degpt, revisiting, quantifying, anythinglen} and allows us to directly compare against the current state of the art for vocabulary-based type inference, STRIDE~\cite{stride}. 
DIRT comprises binaries compiled from open source C/C++ projects with groundtruth type annotations extracted from debug symbols.
The original split consists of 75,656 binaries (738,158 functions) in the train set, and 9,457 binaries (ca. 142,000 functions) in the validation and test set each.
For training optimization experiments, we sample additional binaries from the raw unprocessed DIRT dataset, made available by the authors~\cite{dirtraw}.
Additionally, we use a combination of popular real-world binaries to evaluate struct recovery performance: \texttt{coreutils}, \texttt{wget}, \texttt{grep}, \texttt{gzip} and \texttt{lighttpd}. The binaries used are the exact same versions as in the evaluation of HyRES~\cite{hyres-dataset}. 
We use the IDA-based scripts provided by DIRTY~\cite{dirtygh} to preprocess the binaries and extract them in a format compatible \sysname.
All aforementioned datasets consist of binaries compiled for the x86\_64 architecture. As all compared systems operate on decompiled code or, in the case of HyRES, on the LLVM intermediate representation~\cite{llvm}, we consider the approaches to be architecture-agnostic. 
For our function recovery case study, we use a dataset of ARM embedded firmware binaries that was used in prior research in firmware emulation~\cite{fuzzware, halucinator, safirefuzz}.

\textbf{Baseline selection.}
We compare against \textbf{DIRTY}~\cite{dirty}, a Transformer baseline that introduced the DIRT dataset, and \textbf{STRIDE}~\cite{stride}, the original n-gram-based approach and state of the art on DIRT, for general type inference.
For struct recovery, we compare against \textbf{HyRES}~\cite{hyres}, a hybrid approach specializing on struct recovery, and \textbf{TypeForge}~\cite{typeforge}, an LLM-assisted synthesis-selection method with the same focus, for which we replicate results using the provided open source implementation.
% We do not include OSPREY in the main experiments because HyRES and TypeForge strictly improve over it in both recovery quality and runtime in the original publications; we therefore retain OSPREY as historical context in related work and Table~\ref{tab:relwork-comparison}.

\textbf{Metrics.}
For comparative means, we report $(1)$ overall type accuracy as the percentage of variables with correct fully qualified type, $(2)$ struct type accuracy as accuracy on variables with a struct groundtruth, $(3)$ in-training vs. out-of-training performance on functions seen/unseen during training, $(4)$ precision, recall, F1 as per their classical definitions.

\textbf{Hardware.}
Unless specified otherwise, experiments were conducted on macOS 15 with an M3 Max CPU with 12 performance cores and 128 GB RAM.

\begin{table*}[t]
\centering
\caption{Hyperparameter tuning of training set size and database composition on validation set. Values in brackets indicate n-gram sizes for the databases, 16 DB denotes the original configuration.}
\label{tab:ablations}
\small
\begin{tabular}{lccccc}
\toprule
\textbf{Configuration} & \textbf{Training Size} & \textbf{Memory (GB)} & \textbf{Overall} & \textbf{Out-Train} & \textbf{Throughput} \\
 & &  & \textbf{Accuracy \%} & \textbf{Accuracy \%} & \textbf{ms/func} \\
\midrule
% STRIDE (16 DB) & 75,000 & 15 & 85.11 & 65.9\% \\
% 100k @ [96, 32, 16, 12, 8, 6, 5, 4, 3, 2] & 100,000 & 13 & 86.48 & 66.03 \\
% 150k @ [64, 32, 16, 8, 2] & 150,000 & 13 & 87.90 & 66.76 \\
% 150k @ 16 DB & 150,000 & 27 & 88.53 & 70.94 \\
% 180k @ [64, 32, 16, 8, 5, 3, 2] & 180,000 & 15 & 88.17 & 67.12 \\
% 250k @ [64, 24, 8, 2] & 250,000 & 14 & 89.65 & 67.80 \\
% 250k @ [64, 32, 16, 4] & 250,000 & 19 & 90.19 & 70.04 \\
% 300k @ [48, 16, 8, 2] & 300,000 & 13 & 89.66 & 69.96 \\
% 300k @ [64, 16, 8, 2] & 300,000 & 15 & 90.15 & 71.48 \\
% 420k @ [48, 16, 8, 2] & 420,000 & 16 & 89.37 & 70.88 \\
STRIDE (16 DB) & 75,000 & 15 & 73.75 & 59.87 & 0.13\\
100k @ [96, 32, 16, 12, 8, 6, 5, 4, 3, 2] & 100,000 & 11 & 79.26 & 62.22 & 0.11 \\
150k @ [64, 32, 16, 8, 2] & 150,000 & 12 & 83.11 & 63.90 & 0.07 \\
150k @ 16 DB & 150,000 & 27 & 84.03 & 66.14 & 0.28 \\
180k @ [64, 32, 16, 8, 5, 3, 2] & 180,000 & 15 & 84.63 & 65.09 & 0.08 \\
250k @ [64, 24, 8, 2] & 250,000 & 14 & 86.51 & 65.46 & 0.05 \\
250k @ [64, 32, 16, 4] & 250,000 & 18 & 86.43 & 65.24 & 0.07 \\
300k @ [64, 16, 8, 2] & 300,000 & 15 & 87.99 & 67.23 & 0.05 \\
300k @ [48, 12, 8, 4, 2] & 300,000 & 14 & 88.19 & 67.97 & 0.05 \\
400k @ [48, 12, 8, 2] & 400,000 & 16 & 88.96 & 67.27 & 0.06 \\
\bottomrule
\end{tabular}
\end{table*}

\subsection{RQ1: Finding a Pareto-Optimal Training Configuration}
\label{exp:training}
For RQ1, we set out to explore if an optimized training procedure can achieve better results within the same constraints.
We set the goal as follows: find a better performing combination of number of samples in the training data and number and sizes of produced n-gram databases, such that the final databases are no larger than the databases of the original STRIDE implementation (15 GB) and that the model has no additional overhead in lookup time. 
We explore various configurations with different sizes, not strictly restricting ourselves to the 15 GB bound, to get a better understanding of the relationship between sample size, n-gram contexts, and performance.
Evaluating different training configurations on a dedicated validation set with 20,000 binaries that were not contained in the original train or test sets, Table~\ref{tab:ablations} reports accuracy and memory usage.
We also report the throughput in order to give a feeling for performance trade-offs of different database sizes and configurations.

Increasing training set size from 75k to 300k binaries improves overall accuracy from 85.11\% to 90.15\%.
By adjusting the number of databases and their respective token context sizes, we can further minimize memory requirements while covering more unique context samples at the same time.
The configuration with 300,000 samples and five databases [48, 12, 8, 4, 2] achieves the best memory to out-of-train accuracy trade-off on the validation set.
We designate this configuration \sysname, having the same memory requirements as the original STRIDE configuration but better inference speed due to the reduced number of databases.
The extended training corpus brings the number of functions that are in training but also in the validation set to ca. 67.58\% from STRIDE's 61.05\%, but also leads to better context coverage for functions outside the training dataset.

\subsection{General Type Inference on DIRT}

Table~\ref{tab:main-dirt} presents overall results on DIRT for \sysname and selected baselines.
Testing is performed on the original DIRT test, a corpus of 142,193	functions, identical to the one used in previous work~\cite{dirty, stride, hyres}.
\begin{table}[t]
\centering
\caption{Overall type accuracy on DIRT test set.}
\label{tab:main-dirt}
\small
\begin{tabular}{l|rrr}
\toprule
\textbf{System} & \textbf{Overall (\%)} & \textbf{In-Train (\%)} & \textbf{Out-Train (\%)} \\
\midrule
$DIRTY$ & 75.8 & 89.9 & 56.4 \\
$STRIDE_{repl}$ & 85.06 & \textbf{98.53} & 65.51 \\
$XTRIDE$ & \textbf{90.15} & 98.26 & \textbf{68.66} \\
\bottomrule
\end{tabular}
\end{table}
We report overall type accuracy as the percentage of variables with correct fully qualified type as our main metric in accordance with previous work~\cite{dirty, stride}, not further differentiating any class bias (\textbf{L2}).
We replicate experiments for STRIDE (original configuration and databases ran in our Rust implementation) and take metrics for DIRTY as reported by the authors~\cite{dirty}.
Although the n-gram based approaches are mostly stable, there are slight differences in inferred labels due to numeric instabilities. 
We report the average score across 5 runs in Table~\ref{tab:main-dirt}.
\sysname achieves 90.15\% accuracy with the 5-database configuration (n $\in \{2, 4, 8, 12, 48\}$), improving over STRIDE by 5.09 percentage points statistically significant with the Mann-Whitney U test ($p = 0.040$).
Our Rust implementation processes the DIRT test set at $0.04$ ms per function, where the original Python STRIDE implementation is reported to take $8.2$ ms~\cite{stride} and DIRTY takes between $200$ (low beam setting on GPU) and $8500$ ms (high beam setting in CPU), resulting in a $200$ to $100,000\times$ speed-up. 
While the large jump in overall accuracy is largely due to more n-grams being indexed during training, we also achieve better out-of-train accuracy with an improved accuracy of 68.66\% on functions that are not in the training set as opposed to STRIDE's 65.5\%, improving on \textbf{L4}.
Moreover, \sysname achieves 94.88\% precision in recovering struct types vs. DIRTY's 68.6\%, although their approach explicitly considers and focuses on these with their data layout encoder.
A significant share (15.68\%) of all unidentified structs had a groundtruth label outside of \sysname's vocabulary. 

\begin{table*}[t]
  \centering
  \caption{Struct identification and layout recovery results. Metrics are a macro-average over all benchmark binaries.}
  \label{tab:main-struct}
  \begin{tabular}{l|ccc|ccc|c}
    \toprule
    \textbf{System} & \multicolumn{3}{c|}{\textbf{Struct Identification}} & \multicolumn{3}{c|}{\textbf{Layout Recovery}} & \textbf{Full-Match} \\
     & P & R & F1 & P & R & F1 & \textbf{Accuracy} \\
    \midrule
    HyRES         & 0.958 & 0.834 & 0.891 & 0.892 & 0.717 & 0.795 & $-$   \\
    Typeforge     & 0.738 & 0.456 & 0.556 & 0.663 & 0.507 & 0.555 & 0.201   \\
    XTRIDE\       & 0.512 & 0.251 & 0.329 & 0.738 & 0.828 & 0.768 & 0.742   \\
    $XTRIDE_{PLUS}$ & 0.833 & 0.783 & 0.807 & 0.928 & 0.961 & 0.944 & 0.943 \\
    \bottomrule
  \end{tabular}
\end{table*}

\subsection{RQ2: Comparison to the State of the Art in Struct Recovery}

As undifferentiated accuracy might paint an incomplete picture (\textbf{L2}), we further evaluate our type inference system regarding the recovery of struct types in real-world binaries.
These complex, often user-defined types arguably represent the most interesting class of results, as correct recovery adds the most semantic information and readability to decompiler outputs~\cite{R2I}.

HyRES~\cite{hyres} and TypeForge~\cite{typeforge} are two recently published type recovery approaches with different angles, but both focusing on struct recovery, achieving state-of-the-art results. 
For HyRES, we present the evaluation metrics as originally reported by the authors~\cite{hyres}.
Replicating the results was not possible due to the extreme amounts of required resources, as the original evaluation was run on a machine with 128 cores and 2TB RAM.
To ensure comparability, we run this set of experiments on a dataset being comprised of the exact binaries that were used for HyRES~\cite{hyres-dataset}.
We replicate TypeForge results on this dataset using the provided open-source implementation~\cite{typeforge-github}.
During the evaluation, we encountered multiple challenges and inconsistencies.
For one, the open source implementation only provides scripts to extract groundtruth and to run inference but not to evaluate these two against each other.
Upon request, the authors were not able to provide the original scripts or raw data for verification purposes.
Thus, we implemented evaluation to the best of our abilities.
We plan to make our implementations publicly available to allow for further verification and comparison.
Secondly, the methodology described in their work seems to differ in details from the provided implementation.
For example, they defined \textit{composite types} as "structs and pointers to structs, unions and pointers to unions, and arrays"~\cite{typeforge}.
Yet, their inference implementation only considers variables with observable struct field access patterns as these are used as inputs to the constraint generation.
This rules out arrays as a class of inferrable type.
Finally, we identified and fixed a non-diverging path in the layout propagator, leading to non-terminating loops in some cases (e.g., on our versions of \texttt{wget2} and \texttt{lighttpd}).
Although these limitations contribute to the gap in performance between author-reported results and our replication, upon closer inspection, we are confident that our evaluation fairly reflects the technique's capabilities.

For this task, we trained \sysname in a second configuration to explore upper bound capabilities:
The configuration features a large 16-database configuration and extended the 300k training set by sampling 25\% of the functions in the benchmark binaries.
We denote this configuration $\sysname_{PLUS}$.
We present this configuration explicitly as an upper-bound, in-environment adaptation setting.
It may be seen as a threat to validity when comparing to other systems, but prior work also had significant train-test overlap~\cite{dirty, stride}, and this setup reflects the intended usage of n-gram-based techniques, which are inherently data-dependent and not designed to generalize robustly across substantially different distributions.
In the optimal case, the model is deployed in an at least partially known environment, allowing it to correctly identify the types that are unique to the dataset.
Train-Test overlap ranges between $40.9\%$ (\texttt{lighttpd} O0) and $71.7\%$ (\texttt{coreutils} O3) with a macro average of $61.0\%$.
The total size of the n-gram databases is 48 GB.
To mimic the typical symbol-less reverse engineering scenario and to avoid overfitting on semantics, we manually stripped all occurrences of non-generic variable and function names from the training corpus. The original DIRTY dataset includes non-stripped binaries.
We report metrics on the following tasks in Table~\ref{tab:main-struct}:

\textbf{Struct identification.}
For this, we report the fraction of all struct and pointer-to-struct types identified as structs.
Differences in performance with regard to optimization level are reported in Table~\ref{tab:opt}.
In accordance with TypeForge's implementation and also in alignment with HyRES's methodology, we only consider structs and not arrays or unions.
Our results on replicating TypeForge show significantly lower recall than reported in the original work (77 - 82\%).
We attribute much of this gap to evaluation differences, as TypeForge does not attempt inference for many variables with struct groundtruth if they lack observable struct field access patterns. We count these cases as misses.
\sysname in its base configuration without grounding in the test set has notable difficulties correctly identifying struct types.
As \sysname's inference result is a real, fully-qualified type extracted from real-world binaries, recognition of the correct type always means we also get full type and field names.
TypeForge has no mechanism to recover the semantics of a type and its fields.
HyRES attempts to recover type and field names post-hoc with an LLM. 
They report to correctly recover 67.7\% of field names, where a cosine similarity of the token embeddings $>0.85$ is considered correct.
HyRES excels in identifying structs with a precision and recall of $0.96$ and $0.83$ respectively, finding even those cases that do not emit obvious struct field access patterns due to its inter-procedural analysis.
The fine-tuned $\sysname_{PLUS}$ is strong in correctly identifying structs and can surpass our reproduced TypeForge, being on-par with the F1 score of $0.817$ reported in the original work~\cite{typeforge} as well. 

\textbf{Layout recovery.}
Both HyRES and TypeForge define a field-level metric. 
True positives are correct field offsets, false negatives are missed field offsets, and false positives are incorrect offsets.
For the layout recovery metric, a result is counted as a true positive only if both the offset and the inferred width per offset are correct. 
For \sysname, we compute $P$, $R$ and $F1$ by deconstructing the inferred type class label into its field definitions.
In addition, we report the fraction of fully recovered structs for \sysname and TypeForge, i.e., recovered structs whose layout matches the groundtruth. 
The necessary data to compute this metric was not available for HyRES.
Notably, \sysname performs better on this metric than the baselines, presumably due to its closed vocabulary.
Producing the correct label during inference automatically yields a fully-correct layout grounded in an existing type from the training corpus.
Typical exceptions are different versions of the same struct or minor type confusions after software revisions.
TypeForge's macro-average is heavily influenced by a bad performance on the \texttt{lighttpd} binary, achieving only 0.30 and 0.41 precision and recall respectively.
The system's performance on the other benchmarks is considerably better with up to $0.868$ precision on \texttt{wget}.
Even without fine-tuning, \sysname in its base configuration achieves an F1 of $0.77$ for layout recovery.
The model shows an inverse precision-recall trade-off to HyRES.
With the version with partial knowledge of the test distribution, we achieve even better layout fidelity than HyRES with an F1 score of $0.944$.

\begin{table}[t]
\caption{Struct identification performance of $\text{XTRIDE}_{PLUS}$ on binaries of varying optimization levels across \texttt{coreutils}, \texttt{wget}, \texttt{grep}, \texttt{gzip} and \texttt{lighttpd} (macro-average).}
\label{tab:opt}
\centering
\small
\begin{tabular}{l|ccc}
\toprule
\textbf{Opt-Level} & \textbf{Precision} & \textbf{Recall} & \textbf{F1} \\
\midrule
O0 & 0.878 & 0.757 & 0.811 \\
O1 & 0.901 & 0.815 & 0.855 \\
O2 & 0.838 & 0.811 & 0.824 \\
O3 & 0.724 & 0.774 & 0.747 \\
\bottomrule
\end{tabular}
\end{table}

\textbf{Runtime comparison.}
\label{runtime}
We evaluate the inference speed of the different systems on our real-world benchmark binaries.
For HyRES, we again use the numbers as they appear in the original publication and acknowledge the difference in evaluation environments.
We report runtime of the static dataflow and datalog analysis, excluding runtime of the LLM passes which introduce a further $1.8\times$ on average.
For \texttt{coreutils}, the reported runtime is a per-binary average, the dataset consists of 117 unique binaries.
The TypeForge implementation does not support multi-processing and is thus run sequentially.
We repeated all runs three times and report numbers averaged per project and across optimization levels.
The analysis time for \texttt{lighttpd} diverges notably from the 84 seconds reported in the original work. 
This implementation had issues with the infinite loops that we fixed, possibly exposing other characteristics that led to slow divergence.
For the readability-based refinement via LLM call, we use gpt5-nano instead of gpt4.1-mini. This slows per-binary runtime but had similar practical throughput due to higher API rate limits in a 25-binary subset from \texttt{coreutils}.
In that configuration, the readability-based LLM constraint refinement takes another 78 (\texttt{grep} O0) to 755 (\texttt{lighttpd} O3) seconds. 
Full results are reported in Table~\ref{tab:runtime}.

$\sysname_{PLUS}$ processes ca. 3000 functions per second on average, i.e., 0.33 ms per function.
HyRES and TypeForge do not report per-function timing but per-binary.
Search time increases as the number of n-gram contexts stored in our databases grows (16 vs. 4 databases).
Even in larger configurations, our system is orders of magnitudes faster than the other approaches, showing a $70\times$ to $2,300\times$ speed-up.
Including LLM passes increases this to at least a $180\times$ speed-up against TypeForge.
For semantic recovery, HyRES's approach not only leaves a lot of room for LLM-induced hallucinations but also adds a $1.8\times$ overhead on top of their static analysis passes on average.
While full evaluation of function signature recovery is out of scope of this work, we note that a closed vocabulary approach provides full semantic information at no additional cost upon correct classification.

\begin{table}[t]
\centering
\caption{Runtime performance in seconds, median of ten runs. Numbers are reported per binary.}
\label{tab:runtime}
\small
\begin{tabular}{l|cccc}
\toprule
\textbf{Program} & \textbf{HyRES} & \textbf{TypeForge} & \textbf{\sysname} & \textbf{$\sysname_{PLUS}$} \\
\midrule
wget & 719.07 & 21.38 & 0.13 & 0.31 \\
lighttpd & 498.91 & 240.75 & 0.36 & 0.89 \\
grep & 293.45 & 63.47 & 0.19 & 0.47 \\
gzip & 148.04 & 24.86 & 0.13 & 0.33 \\
coreutils & 189.32 & 9.2 & 0.05 & 0.08 \\
\bottomrule
\end{tabular}
\end{table}

\subsection{RQ3: Threshold Analysis}

For a fixed threshold $t$, let $K_t$ be the number of kept (non-abstained) predictions and $C_t$ the number of exactly correct kept predictions.
The selective exact accuracy at threshold $t$ is defined as
\[
\text{sel-acc}_{\text{exact}}(t) = \frac{C_t}{K_t}.
\]
The corresponding variable risk is
\[
\text{var-risk}_{\text{exact}}(t) 
= 1 - \text{sel-acc}_{\text{exact}}(t)
= 1 - \frac{C_t}{K_t}
= \frac{K_t - C_t}{K_t}.
\]
That is, $\text{var-risk}_{\text{exact}}(t)$ is the error rate on the non-abstained subset.
Accordingly, $\text{struct-risk}_{\text{exact}}(t)$ is the risk only considering variables with a struct type groundtruth.
Figure~\ref{tab:thresholds} reports performance at different confidence thresholds for the main \sysname model configuration on the stripped DIRTY test set. 

We show that the procedure behaves intuitively, allowing for fine-tuning of the threshold depending on the task and context.
Increasing the threshold can reduce the error rate significantly at the cost of providing type predictions for fewer variables.
The other approaches offer no quantitative measure of reliability allowing to discard low-confidence predictions.
If the test distribution is known a priori, we recommend full calibration on an in-domain validation set as described in~\ref{scoring}.
If the distribution remains stable, calibration can be reused across deployments with similar data.

% \begin{table}[t]
% \small
% \begin{tabular}{l|ccc}
% \toprule
% Threshold & Coverage \% & $\text{var-risk}_{exact}$ & $\text{struct-risk}_{exact}$ \\
% \midrule
% none & 100.0 & 0.153 & 0.089 \\
% 0.40 & 90.75 & 0.088 & 0.058 \\
% 0.65 & 77.03 & 0.036 & 0.030 \\
% 0.90 & 47.84 & 0.024 & 0.022 \\
% \bottomrule
% \end{tabular}
% \end{table}

\begin{figure}[t]
\centering
\small
\begin{tikzpicture}
% Define color-blind-friendly colors (from the Okabe-Ito palette)
\definecolor{cbOrange}{HTML}{D55E00}
\definecolor{cbBlue}{HTML}{0072B2}

\pgfplotsset{every axis/.append style={font=\small}}

% Left axis: Coverage
\begin{axis}[
  name=covaxis,
  width=\dimexpr\columnwidth-40pt\relax, 
  height=0.62\columnwidth,
  xlabel={Confidence threshold},
  ylabel={Coverage (\%)},
  symbolic x coords={none,0.40,0.65,0.90},
  xtick=data,
  enlarge x limits=0.12,
  ymin=40, ymax=105,
  grid=both,
  % Legend in the top-right, stacked vertically
  legend style={
    at={(0.97,0.97)},
    anchor=north east,
    legend columns=1,
    draw=black,
    fill=white,
    fill opacity=0.95,
    text opacity=1,
    font=\tiny,
    row sep=1pt,
  },
  legend cell align={left},
]

% Coverage (Now using cbOrange)
\addplot+[cbOrange, thick, mark=*, mark options={solid}, mark size=2.4pt]
coordinates {
  (none,100.0)
  (0.40,90.75)
  (0.65,77.03)
  (0.90,47.84)
};
\addlegendentry{Coverage (\%)}

% Legend entries for risk series (Now using cbBlue)
\addlegendimage{cbBlue, thick, mark=square*, mark options={solid}, mark size=2.4pt}
\addlegendentry{$\mathrm{var\text{-}risk}_{\mathrm{exact}}$}

\addlegendimage{cbBlue, thick, mark=triangle*, mark options={solid}, mark size=2.6pt}
\addlegendentry{$\mathrm{struct\text{-}risk}_{\mathrm{exact}}$}

\end{axis}

% Right axis: Risk
\begin{axis}[
  at={(covaxis.south west)},
  anchor=south west,
  width=\dimexpr\columnwidth-40pt\relax, 
  height=0.62\columnwidth,
  symbolic x coords={none,0.40,0.65,0.90},
  xtick=\empty,
  axis x line=none,
  axis y line*=right,
  ylabel={Risk},
  ymin=0, ymax=0.16,
  ytick={0,0.05,0.10,0.15},
  yticklabels={$0$,$0.05$,$0.1$,$0.15$},
  tick align=inside,
  y tick label style={xshift=-2pt, font=\small},
  ylabel style={xshift=-2pt, font=\small},
]

% var-risk_exact (Now using cbBlue)
\addplot+[cbBlue, thick, mark=square*, mark options={solid}, mark size=2.4pt, forget plot]
coordinates {
  (none,0.153)
  (0.40,0.088)
  (0.65,0.036)
  (0.90,0.024)
};

% struct-risk_exact (Now using cbBlue)
\addplot+[cbBlue, thick, mark=triangle*, mark options={solid}, mark size=2.6pt, forget plot]
coordinates {
  (none,0.089)
  (0.40,0.058)
  (0.65,0.030)
  (0.90,0.022)
};

\end{axis}
\end{tikzpicture}
\caption{Coverage-Risk trade-offs at different confidence thresholds on the stripped version of the DIRTY benchmark.}
\label{tab:thresholds}
\end{figure}

\subsection{RQ4: Function Signature Recovery}
\label{subsec:eval-funcname}

For the function signature recovery task as described in~\ref{funcrec}, calls to functions are treated like variables.
The surrounding tokens are extracted and matched against n-gram databases, the class label inferred from a separate function vocabulary.
On the stripped version of the DIRT test set, $\sysname_{PLUS}$ correctly identifies $78.97\%$ of all function calls.
This result is heavily biased by the top few percent of frequently occurring functions.
Common library functions from popular open-source projects, such as \texttt{av\_log} from \texttt{ffmpeg} with 1.01\%, are the most frequently correctly inferred functions.

As performance appears numerically low, we conduct a case study to get a better feeling for its impact in real-world applications: we mimic a typical reverse engineering workflow where we need to identify key functions in the complete absence of symbols, in this case on a set of embedded firmware binaries (cf.~\ref{setup}).
Hardware Abstraction Layer (HAL) functions often interact with peripherals via MMIO and are of special interest during emulation, rehosting and fuzzing of embedded firmware as they require special handling.
They are commonly hooked in rehosting systems like HALucinator~\cite{halucinator}.
To enable the model to recognize the desired HAL functions, we add two binaries to the training corpus: \texttt{p2im\_plc}, another common benchmark binary for embedded rehosting~\cite{p2im, halucinator, safirefuzz}, and a test project compiled to include as many HAL functions as possible.
The first provides real-world usage contexts while the latter maximizes vocabulary coverage of HAL functions. Both these binaries, as well as our benchmark binaries, use common functions from the STM32 HAL library. 
We do not add any code from the binaries we evaluate on to the training corpus, thus the percentage of test-train overlap lies in the $1.5$ - $4.5\%$ range only.
Table~\ref{tab:fname} reports precision, recall, and F1 for identifying HAL functions at threshold $0.4$.

\begin{table}[t]
\centering
\caption{Function signature recovery results on embedded firmware samples at confidence threshold 0.4.}
\small
\begin{tabular}{l|ccc}
\toprule
\textbf{Sample} & \textbf{Precision} & \textbf{Recall} & \textbf{F1} \\
\midrule
st\_plc & 0.604 & 0.441 & 0.510 \\
drone & 0.567 & 0.283 & 0.378 \\
tcp\_echo & 0.365 & 0.409 & 0.386 \\
\midrule
Average & 0.515 & 0.374 & 0.431 \\
\bottomrule
\end{tabular}
\label{tab:fname}
\end{table}

Function signature recovery achieves moderate precision but low recall, indicating that the approach is conservative and that many usage patterns do not match contexts in the training corpus.
Our function recovery mechanism aggregates labels and produces one final inference result per global function start address, so repeated calls inside a binary are not double-counted at the reporting stage.
For groundtruth selection, we consider all functions with names starting with \texttt{HAL\_}. 
The \texttt{st\_plc} binary contains 1127 inferrable functions, 286 of which are HAL functions.
Our approach correctly identifies 126 HAL functions with a precision of $60.39\%$, while 21 HAL functions were missing from the vocabulary, accounting for $13.04\%$ of false negatives.
On \texttt{tcp\_echo}, only 69 of 732 functions are HAL functions, resulting in lower precision.
Overall, we correctly identify $44.06$, $28.33$, and $40.91\%$ of all HAL functions in the binaries for \texttt{st\_plc}, \texttt{drone} and \texttt{tcp\_echo} respectively.

We consider this result promising, demonstrating utility for identifying likely HAL functions.
Especially in self-contained binary blobs, like many embedded firmware are, that contain hundreds of functions, this extremely low-overhead approach is a fast way to condense the number of functions to manually look at when reverse engineering.

% \subsection{Case Study: Applying Types in Ghidra}

% We demonstrate \sysname's utility by applying inferred types to a decompiled function in Ghidra.
% Figure~\ref{fig:case-study} shows before-and-after decompilation output.

% \todo{Create case study figure as gfx/case-study.pdf}

% Before type application, the decompiled code contains generic types (\texttt{undefined4}, pointer casts) and pointer arithmetic.
% After applying \sysname's inferred types with confidence $\geq 0.95$, struct field accesses are resolved, variable names are improved, and the code is substantially more readable.

%!TEX root = ../paper.tex
\section{Discussion}
\label{sec:discussion}

Recovering types for stripped binaries is only useful when the recovered information survives contact with downstream tooling. 
The results indicate that struct recovery is constrained as much by semantic fidelity and integration requirements as by raw layout reconstruction. 
Grounding predictions in a closed vocabulary of fully qualified, real-world types yields outputs that are immediately actionable. 
A correct prediction implies not only a compatible layout but also a stable semantic identifier that can be propagated through decompiler types, signatures, and dataflow views without an additional naming step.
 
Our results also indicate that improvements are not primarily tied to more elaborate inference, but to how sparsity is managed at training and indexing time. 
Scaling the corpus while reducing the number of n-gram databases improves test accuracy without expanding the memory footprint, suggesting that database composition is a more effective lever than using many overlapping context indices. 
This behavior is consistent with the intuition that diverse, well-chosen context lengths improve coverage of predictive neighborhoods in token space, whereas additional databases can contribute redundant evidence with diminishing returns.

For composite structs, fully correct layouts follow at a high rate if a struct is correctly identified, reflecting the advantage of predicting fully-qualified grounded types. 
The dominant failures arise earlier, at identification, particularly for types absent from the vocabulary. 
This contrasts with layout synthesis approaches such as HyRES and TypeForge, which infer shapes from program evidence.
Grounded prediction offers high semantic fidelity when the ecosystem contains recurring types, while synthesis remains useful when types are novel or definitions are unavailable.

The confidence analysis further bridges benchmark accuracy and deployability. 
The precision–recall trade-offs obtained via thresholding indicate that the model supports a selective mode in which predictions can be applied automatically at high confidence while abstaining on ambiguous cases. 
This is particularly relevant for automated pipelines, where even a small rate of incorrect type application can cascade into misleading decompiler output. 

Finally, the embedded firmware case study extends the same usage-based principle to function signature recovery. 
The observed profile, high precision with modest recall, is consistent with a conservative similarity signal that is useful for triage.
Surfacing a small set of high-interest, in this case library or HAL functions can help to anchor manual analysis.
In throughput-bound settings, a fast abstaining predictor can prioritize analyst attention without broad noisy labeling.

The main limitation of the n-gram-based approach follows directly from the design choice that enables semantic fidelity, as predictions cannot exceed the closed vocabulary. 
Misses on out-of-vocabulary types are therefore expected, and improvements in corpus breadth mainly reduce this boundary rather than remove it. 
Accordingly, we do not claim open-world recovery.
The approach is best suited for partially known environments with recurring real types, such as library-rich software stacks, firmware ecosystems, or supply-chain settings with substantial component reuse.
The results suggest a natural division of labor where grounded prediction covers the frequent and recurrent portion of real-world type ecosystems with high fidelity and very low overhead and residual cases can be deferred to heavier reconstruction.

% !TEX root = ../paper.tex
\section{Related Work}
\label{sec:relwork}
In this section, we discuss related work in type recovery.

\textbf{Traditional static analysis approaches.}
Early approaches to type recovery relied on constraint solving and heuristics.
Decompilers such as IDA Pro and Ghidra can recover primitive data types with reasonable accuracy~\cite{ida,ghidra}.
However, these tools struggle significantly with user-defined composite data types, often misidentifying structures as generic integer types.

Constraint-solving methods such as TIE~\cite{tie}, Retypd~\cite{retyped}, and OSPREY~\cite{osprey} employ whole-program memory dependency analysis and heuristic rules to infer types and layouts without debug information.
OSPREY uses probabilistic inference based on memory access patterns and dynamic dataflow, achieving 90\% precision on primitive types on a \texttt{coreutils} benchmark.
For struct layout recovery, OSPREY achieves F1 0.869 on the \texttt{coreutils} dataset, though with substantial computational overhead (averaging 528.24 seconds per binary).
Static inference methods in general tend to be computationally expensive, limiting their scalability and practical speed. 
The BinSub~\cite{binsub} authors acknowledge that existing implementations of type inference systems that are based on subtyping and polymorphism are too inefficient to achieve adoption in production. 
Their work introduces algebraic subtyping to the domain, leading to a faster method for binary type reconstruction, framing the work as an alternative approach to the static methods in mainstream decompilers rather than as a post-hoc enhancement.

\textbf{Machine learning approaches.}
With the rise of machine learning, data-driven approaches have emerged to augment decompilation.
Especially Transformers and LLMs have been successful recently in type recovery tasks across domains~\cite{dirty, resym, codetidal, empirical, idioms}.
DIRTY~\cite{dirty} is a Transformer-based model that post-processes decompiled files to generate variable names and types.
It supports 48,888 types from open source C code and uses a Data Layout Encoder to incorporate memory layout information.
DIRTY achieves 75.8\% overall type accuracy but performance notably decreases on testing sets due to its closed vocabulary approach.
Furthermore, DIRTY requires over one second per function on GPU and more than $8\times$ that on CPU, making it prohibitive for large-scale analysis.
TYGR~\cite{tygr} uses Graph Neural Networks with intra-procedural dataflow analysis on VEX IR~\cite{valgrind} to infer basic and struct variable types from a closed dictionary.
TYGR introduces the TYDA dataset due to the high occurrence of undesired artifacts in prior datasets~\cite{tygr}.
On the \texttt{coreutils} O0 benchmark, TYGR reports overall accuracy of 74.5\% and struct accuracy of 40.6\%.
While TYGR aims to predict struct shapes and member types, other work reports that it fails to recover member layout comprehensively, with low F1 scores (9.1\%) for layout recovery in comparison evaluations~\cite{typeforge}.
ByteTR~\cite{bytetr} conducts inter-procedural analysis using its BytePA algorithm and uses a Gated GNN for variable-level semantic representation.
It explicitly does not support recovery of full structure layouts or specific structure identifier names, considering this a difficult problem for future work.

\textbf{Hybrid and advanced approaches.}
Recent work combines static analysis with LLMs to improve semantic recovery.
HyRES~\cite{hyres} is a hybrid reasoning technique combining Datalog-based reasoning on LLVM IR~\cite{llvm} with LLMs and heuristics.
The approach performs intra- and inter-procedure dataflow analysis with structure aggregation to overcome incomplete dataflow.
It significantly outperforms OSPREY and ReSym~\cite{resym} in struct pointer identification and layout recovery with substantially lower overhead than OSPREY.
ReSym combines fine-tuned LLMs with a lightweight Prolog-based reasoning system.
It uses Andersen's algorithm for dataflow analysis on decompiled code and aggregates LLM queries to suppress uncertainty.
ReSym substantially outperforms OSPREY and DIRTY in variable name and type recovery.
At the same time, its dataflow analysis is inherently undecidable, leading to incomplete type grouping, and identifying type-agnostic arguments is challenging, reducing precision.
TypeForge~\cite{typeforge} employs a two-stage synthesis-selection strategy.
The first stage uses a Type Flow Graph for efficient hint collection with conflict-aware type propagation to prevent erroneous propagation.
The second stage uses LLM-assisted readability-guided selection to choose the best-fit type declaration.
TypeForge significantly outperforms DIRTY, OSPREY, ReSym, and TYGR in composite data type identification and layout recovery.
%!TEX root = ../paper.tex
\section{Conclusion}
\label{sec:conclusion}

In this work, we position n-gram-based type recovery as a practical component for high-throughput program analysis workflows.
Our system \sysname combines an optimized n-gram database portfolio that improves accuracy without inflating memory, thresholdable confidence, and an experimental extension to function signature recovery.

\sysname achieves 90.15\% overall accuracy and 68.66\% out-of-training accuracy, improving over the state of the art by 5.09 and 3.15 percentage points respectively while reducing the database count and keeping the memory footprint consistent. 
Moreover, our implementation yields $10^3$--$10^5\times$ speedups over static-analysis-heavy, LLM-centered pipelines. 
By thresholding on confidence, users can trade coverage for reliability and avoid applying low-quality types that would otherwise degrade downstream decompiler output.
This places high-fidelity, vocabulary-grounded type annotation within reach of interactive decompilation and large-scale automated analysis.
The function recovery case study suggests that the same n-gram matching principle provides a low-cost similarity signal for auxiliary tasks, presenting the first unification of type and function inference into a single approach to the best of our knowledge.
We achieve moderate precision (51.47\% on average), which supports conservative triage during early binary analysis.
Finally, the struct type recovery results demonstrate that grounded prediction yields strong end-to-end correctness if types exist in the vocabulary.

Overall, the results show that n-gram type recovery can offer an effective efficiency-fidelity trade-off in partially known environments.
With appropriate training scale, database design, calibrated confidence, and in-domain deployment, it becomes a practical component of robust high-throughput program analysis workflows.

\subsection*{Availability}
The source code for the \sysname Rust implementation, as well as the trained databases used for our experiments, will be accessible here: \url{https://github.com/pr0me/XTRIDE}.

\begin{acks}
This research was conducted in part while the first and second authors were employed by Binarly, Inc., which provided support for the work.
\end{acks}

\bibliographystyle{ACM-Reference-Format}
\bibliography{input/mybib}

% \appendix









\end{document}